\documentclass{mn2e}
\usepackage{epsfig}

\usepackage{graphicx}

\newif\ifAMStwofonts
\AMStwofontstrue

\title[ELAIS paper IX: the $90\mu$m luminosity function]
{The European Large Area ISO Survey IX: the 
$90\mu$m luminosity function from the Final Analysis sample
}
\author[Serjeant, et al.]
  {
{Stephen Serjeant$^{1\ast}$, 
Alberto Carrami\~{n}ana$^{2\ast}$, 
Eduardo Gonz\'ales-Solares$^{14}$, 
}\vspace*{0.3cm}\\ {\LARGE\rm
Phillipe H\'{e}raudeau$^{4,5}$, 
Ra\'ul M\'ujica$^{2\ast}$, 
Ismael Perez-Fournon$^6$, 
Nicola Sedgwick$^{1\ast}$, 
}\vspace*{0.3cm}\\ {\LARGE\rm
Michael Rowan-Robinson$^7$, 
Alberto Franceschini$^9$, 
Thomas Babbedge$^7$, 
}\vspace*{0.3cm}\\ {\LARGE\rm
Carlos del Burgo$^{13}$, 
Paolo Ciliegi$^8$, 
Andreas Efstathiou$^{12}$, 
Fabio La Franca$^{10}$, 
}\vspace*{0.3cm}\\ {\LARGE\rm
Carlotta Gruppioni$^8$, 
David Hughes$^2$, 
Carlo Lari$^{11}$,
Seb Oliver$^3$,
Francesca Pozzi$^{11}$, 
}\vspace*{0.3cm}\\ {\LARGE\rm
Manfred Stickel$^5$, 
Mattia Vaccari$^9$
\vspace*{0.2cm}
}\\
$^1$Centre for Astrophysics \& Planetary Science, School of Physical
Sciences, University of Kent, Canterbury, Kent CT2 7NR, UK\\
$^2$Instituto Nacional de Astrof\'{\i}sica, \'Optica y Electr\'onica,
Luis Enrique Erro 1, Tonantzintla, Puebla 72840, M\'exico\\
$^3$Astronomy Centre, CPES, University of Sussex, Falmer, Brighton BN1
9QJ\\
$^4$Kapteyn Astronomical Institute, Postbus 800, 9700 AV Groningen,
The Netherlands\\
$^5$Max-Planck-Institut f\"{u}r Astronomie, K\"{o}ningstuhl 17,
D-69117, Heidelberg, Germany\\
$^6$Instituto de Astrofisica de Canarias, C/V\'\i a L\'actea s/n. 38200 La
Laguna, Tenerife, Spain\\
$^7$Astrophysics Group, Imperial College London, Blackett Laboratory,
Prince Consort Road, London SW7 2BW\\
$^8$Bologna Astronomical Observatory, Via Ranzani 1, 40127 Bologna,
Italy\\
$^9$Dipartimento di Astronomia, Universit\`a di Padova, Vicolo
Osservatorio 5, I-35122 Padova, Italy\\
$^{10}$Dipartimento di Fisica, Universit\`a Roma TRE, Viea della
Vasca Navale, 84 00146, Roma, Italy\\
$^{11}$Istituto do Radioastronomia, Via P. Gobetti, 101, 40129,
Bologna, Italy\\
$^{12}$Cyprus College, 6 Diogenes Street, Engomi, P.O. Box 22006, 1516
Nicosia, Cyprus\\
$^{13}$European Space and Technology Centre (ESTEC), Keplerlaan 1,
2201AZ Noordwijk, The Netherlands\\
$^{14}$Institute of Astronomy, University of Cambridge, Madingley Road,
Cambridge, CB3 0HA, UK\\
$^\ast$Visiting Astronomer, Observatorio Astrof\'{\i}sico Guillermo
Haro, Cananea, Sonora, M\'exico
}
\date{Received 2004}
\pubyear{2004}

\begin{document}


\input BoxedEPS.tex
\SetEPSFDirectory{./}

\SetRokickiEPSFSpecial
\HideDisplacementBoxes

\label{firstpage}

\maketitle

\begin{abstract}
We present the $90\mu$m luminosity function of the Final Analysis of the
European Large Area {\it ISO} Survey (ELAIS), extending the sample size of 
our previous analysis (paper IV) by about a factor of $4$. Our
sample extends to $z=1.1$, $\sim50$ times the comoving volume of
paper IV, and $10^{7.7}<h^{-2}L/L_\odot<10^{12.5}$. 
From our optical spectroscopy campaigns of the 
northern ELAIS $90\mu$m survey ($7.4$ deg$^2$ in total, to
$S_{90\mu\rm m}\ge70$mJy), we obtained redshifts 
for $61\%$ of the sample ($151$ redshifts) 
to $B<21$ identified at $7\mu$m, $15\mu$m, $20$cm
or with bright ($B<18.5$) optical identifications. The selection
function is well-defined, permitting the construction of the 
$90\mu$m luminosity function of the Final Analysis catalogue 
in the ELAIS northern fields, which is in excellent agreement
with our Preliminary Analysis luminosity function in the ELAIS S1
field from paper IV. The luminosity function 
is also in good 
agreement with the IRAS-based prediction of Serjeant \& Harrison
(2004), which if correct requires luminosity evolution of
$(1+z)^{3.4\pm 1.0}$ for consistency with the source
counts. This implies an evolution in comoving volume averaged star
formation rate at $z\stackrel{<}{_\sim}1$ consistent with that derived
from rest-frame optical and ultraviolet surveys. 
\end{abstract}


\section{Introduction}\label{sec:introduction}

Recent infrared surveys have revealed the existence of strongly
evolving populations of far-infrared-luminous galaxies. For example, 
the discovery of sub-mm luminous galaxies (Smail et al. 1997, 
Hughes et al. 1998) with
median redshifts $>2$ (Chapman et al. 2003) proves the existence of a
radical change in the demographics of star formation. Luminous
infrared galaxies ($\sim10^{11}L_\odot$) contributed at least as much 
to the comoving star formation rate at $z\sim2$ as all
optical/ultraviolet identified sources, 
but are negligible
contributors locally. Furthermore, the spectrum of the extragalactic
background (e.g. Puget et al. 1996, Fixen et al. 1998, Lagache et
al. 1999) implies that approximately half the luminous energy emitted
by stars throughout the history of the Universe was absorbed and
re-emitted by dust in the far-infrared. 
Differences between the evolution of infrared-luminous galaxies and
that of galaxies selected in the rest-frame optical/UV may be in part
attributable to the changes in the demographics of obscured
vs. unobscured star formation, but may also be related to large scale
structure density variations in the small-area surveys (e.g. $\ll1$
square degree) conducted to
date (see discussion in Oliver et al. 2000). Consequently, 
while some of the most significant earlier contributions have been in
small-area surveys (e.g. Hughes et al. 1998), the emphasis has since
shifted to wide-area surveys ($\sim1-100$ square degrees) which are
less affected by cosmic 
variance problems (e.g. Oliver et al. 2000, Lonsdale et al. 2003). 

The European Large Area {\it ISO} Survey (ELAIS) was the
largest open time 
survey on the Infrared Space Observatory ({\it ISO}), and resulted in the
delivery of the largest catalogue of any survey on {\it ISO}
(Rowan-Robinson  
et al. 2004) from both the ISOCAM instrument 
(Cesarsky et al. 1996)
and ISOPHOT instrument
(Lemke et al. 1996). 
The science goals are diverse, and discussed in detail
in Oliver et al. (2000, paper I). Following the Preliminary Analysis of the
ELAIS data (Serjeant et al. 2000, paper II, Efstathiou et al. 2000,
paper III), a more
exhaustive analysis of the data has been made resulting in deeper and
better calibrated source lists (Lari et al. 2001, Gruppioni et
al. 2002, H\'{e}raudeau et al. 2004 paper VIII), and is referred to as
the ELAIS Final Analysis. The ELAIS survey complements the deeper
surveys conducted by ISO (e.g. Franceschini et al. 2002 and references
therein; Rodighiero et al. 2003)
and the shallower ISOPHOT Serendipity Survey (Stickel et al. 2004). 

The Final Analysis catalogue paper (Rowan-Robinson et al. 2004)
presents the source list and redshifts obtained to date from ELAIS. 
These sources have been studied in earlier papers in this series. 
For example, Morel et al. (2001, paper VI) present the first
hyperluminous galaxy discovered in ELAIS, several more of which are
catalogued in the Rowan-Robinson et al. (2004) paper; 
Pozzi et al. (2003) compare the $15\mu$m, H$\alpha$ and
$1.4$GHz 
star formation rates of galaxies in the S2 field; 
Vasainen et
al. (2002) report the near-infrared identifications of ELAIS galaxies,
in the process obtaining a improved flux calibration which implies a
modest correction to the Preliminary Analysis source counts (Serjeant
et al. 2000) consistent with the Final Analysis source counts
(Gruppioni et al. 2002); Basilakos et al. (2003, paper VII); 
and Manners et al. (2003a,b) examine the X-ray
properties of ELAIS galaxies. The evolution of AGN in ELAIS is
examined in Matute et al. (2002). 
The ELAIS fields have also been the subject of many subsequent 
multi-wavelength surveys (e.g. Ciliegi et al. 1999, Gruppioni et
al. 1999, Alexander et al. 2001 (paper V), 
Scott et al. 2002, Fox et al. 2002, Manners et al. 2003a,b) making the
fields among the best studied degree-scale areas on the sky. 
Finally, 
the main ELAIS survey areas will shortly be observed by the SWIRE
Legacy Survey (Lonsdale et al. 2003) 
on the {\it Spitzer Space Telescope}, making this
catalogue extremely important for the rapid science exploitation of
the {\it Spitzer} data. 

The multiwavelength galaxy luminosity functions are one of the main
observational constraints on the evolution of dust-enshrouded star
formation in galaxy populations. 
In Serjeant et al. (2002, paper IV) we derived the $90\mu$m luminosity
function of ELAIS galaxies in the Preliminary Analysis of the 
ELAIS S1 field (Efstathiou et al. 2000). We found good
agreement with local determinations at the low redshift end, and 
evidence for pure luminosity evolution at the rate of
$(1+z)^{2.45\pm0.85}$. This is consistent with the optically-derived
evolution in the comoving volume-averaged star formation rate
(e.g. Glazebrook et al. 2003). In this
paper we present the $90\mu$m luminosity function for the Final
Analysis of the remaining areas (H\'{e}raudeau et al. 2004, paper VIII). 
Section \ref{sec:method} describes the sample selection
and the data acquisition, and section \ref{sec:results} presents 
the $1/V_{\rm max}$ luminosity
function. 
Section \ref{sec:discussion} discusses 
our results in the
context of source count model evolution and the comoving star
formation history. 

We assume a Hubble constant of $H_0=100h=72$ km s$^{-1}$
Mpc$^{-1}$, and a cosmology of $\Omega_{\rm M}=0.3$,
$\Omega_\Lambda=0.7$ throughout this paper. We adopt the convention of
converting from
$90\mu$m monochromatic luminosities to bolometric luminosities
assuming $\nu L_\nu=$constant, 
i.e. $10^{10}L_\odot$ corresponds to 
approximately $1.159\times10^{24}$ W Hz$^{-1}$ at
$90\mu$m. This choice simplifies the comparisons with other
studies and does not affect the results in this paper. 

\section{Method}\label{sec:method}
\subsection{Sample selection}
Our principal sample is the ELAIS $90\mu$m Final Analysis sample of 
H\'{e}raudeau et al. (2004) in the Northern ELAIS fields, covering
a total of $7.4$ square degrees to a flux limit of $70$mJy. The
completeness as a function of flux is given in H\'{e}raudeau et
al. (2004), and varies from $\sim100\%$ at $150$mJy to $\sim50\%$ 
at $70$mJy. 
The source counts at the faintest end (e.g. $70-130\mu$Jy)
differ from those of the Preliminary Analysis (paper II, paper
IV). These data points were neglected in the fit to the source counts,
from which the evolution rate was derived, so the evolution quoted in
paper IV is unaffected by the change in the catalogues at the faintest
end. Notably, the source counts of the {\it optically identified}
sources are identical between the Preliminary and Final analyses. We do
not have Final Analysis catalogues for the small ELAIS survey areas
(paper I),
so we supplement the Final Analysis catalogue with optically identified
sources from the Preliminary Analysis, which are almost
entirely in these supplementary small fields. 
The exclusion of these 
sources does not affect determined luminosity function, apart from
decreasing the spectroscopic sample size by seven objects. The
catalogue has a total of $420$ 
galaxies, of which we have optically identified $249$ to magnitude
limits discussed below. We have obtained optical spectroscopy of $151$
of these optically identified galaxies to date.  

\begin{figure}
\centering
\ForceWidth{5in}
\hSlide{-1.8cm}
\vspace*{-1cm}
\BoxedEPSF{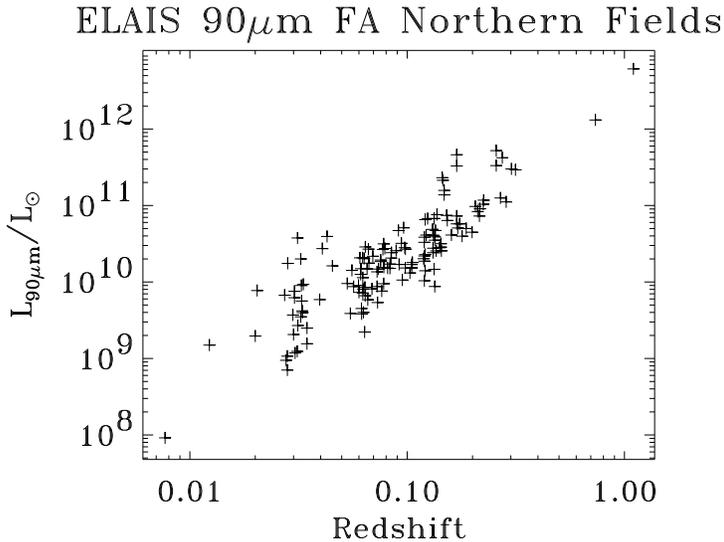}
\caption{\label{fig:luminosity_redshift}
Luminosity-redshift plane for spectroscopically identified ELAIS
$90\mu$m Final Analysis galaxies from the northern ELAIS
fields. Monochromatic luminosities at $90\mu$m have been converted to
bolometric luminosities assuming $\nu L_\nu$=constant.
No correction has been made for pure luminosity evolution. 
} 
\end{figure}

Figure \ref{fig:luminosity_redshift} plots the luminosity-redshift
plane for our sample. 
For the purposes of K-corrections, the 
rest-frame galaxy spectrum in the vicinity of $90\mu$m
is assumed to follow $d\log L_\nu/d\log\nu=-2$, though this choice has
very little affect on the derived luminosity function. 
There are no photometric redshifts used in the derivation of the
luminosity function in this paper. 

\begin{figure}
\centering
\ForceWidth{5in}
\hSlide{-1.8cm}
\vspace*{-1cm}
\BoxedEPSF{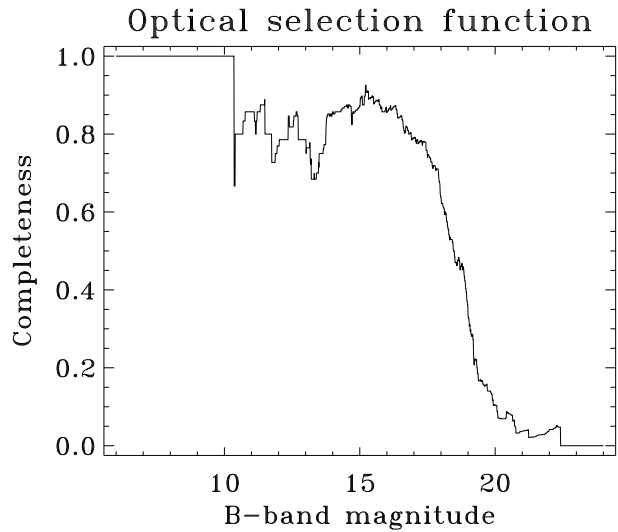}
\caption{\label{fig:optical_selection_function}
Completeness of the optical spectroscopy, as a function of apparent
$B$ magnitude. At each magnitude $m$, the spectroscopic 
completeness plotted refers
to galaxies with magnitudes in the interval $m-1$ to
$m+1$. This curve has structure on scales smaller than $\pm1$
magnitude (e.g. discrete changes) as individual objects move in and
out of the bin being considered. 
}
\end{figure}

\subsection{Data acquisition}
ELAIS $90\mu$m galaxies were optically
identified either purely from the Digitised Sky Survey (for which
paper IV found $B<18.5$ is the limit for reliable identifications), or
from the abundant multi-wavelength data from the ISOCAM
mid-infrared ELAIS data (paper II, Gruppioni et al. 2003) and/or the
VLA follow-up mapping (Ciliegi et al. 1999). For the $90\mu$m galaxies
identified with ISOCAM or the VLA, optical identifications brighter
than $B<21$ were sought using the Digitised Sky Survey and our own
optical imaging (Rowan-Robinson et al. 2004). Given the astrometric
errors of the multiwavelength identifications ($\sim1''-3''$) this
magnitude limit is much more than sufficiently bright to ensure
reliable optical identifications. 

The optical identifications themselves are presented and 
discussed in Rowan-Robinson
et al. (2004). It is possible that despite the careful analysis in
that paper, some of the identifications may be wrong; nevertheless,
the almost-universal appearance of emission lines in our spectra (see
below) implies that the overwhelming majority
of the identifications {\it must} be correct. 
Even in cases where the identification is wrong, the spurious
identification will often have the same redshift as the true
identification since star-forming galaxies well-known to be more
likely to have close companions than the field galaxy population as a
whole (e.g. Surace et al. 2004). Similarly, it is possible that at the
$\sim40''$ resolution of 
ISO, some point sources will in fact be blends of multiple
sources. However, the ELAIS $90\mu$m survey is far from the point
source confusion limit (e.g. Rowan-Robinson 2001), so any blends are
likely to be physically associated galaxy pairs, such as interacting
or merging systems. Again, this would not lead to incorrect redshift
estimates for the purposes of the luminosity function. Also, in local
galaxies there is little evidence that galaxies in pairs each
contribute comparably to their total unresolved far-infrared flux 
(e.g. Surace et al. 2004). 
In summary, for the purposes of this paper we will assume that the
identifications are all correct. 

Spectroscopic campaigns were made on several runs 
at the William Herschel Telescope; 
the Guillermo-Haro $2.2$m, Cananea, Mexico; and the Kitt Peak WIYN facility. 
The Kitt Peak and William Herschel data is described in Perez-Fournon
et al. (in preparation). The Guillermo-Haro data was taken on
17-30 June 2001, 6-13 May 2002 and 1-10 July 2002 using the LFOSC and
Boller \& Chivens low dispersion 
spectrographs. Redshift errors are dominated by wavelength
calibration accuracy ($\sim1-2$\AA) and are negligible for the purposes
of this paper. In total, we
obtained spectroscopic redshifts of $61\%$  of the sample, listed in
Rowan-Robinson et al. (2004). 

\begin{figure}
\centering
\ForceWidth{5in}
\hSlide{-1.8cm}
\vspace*{-1cm}
\BoxedEPSF{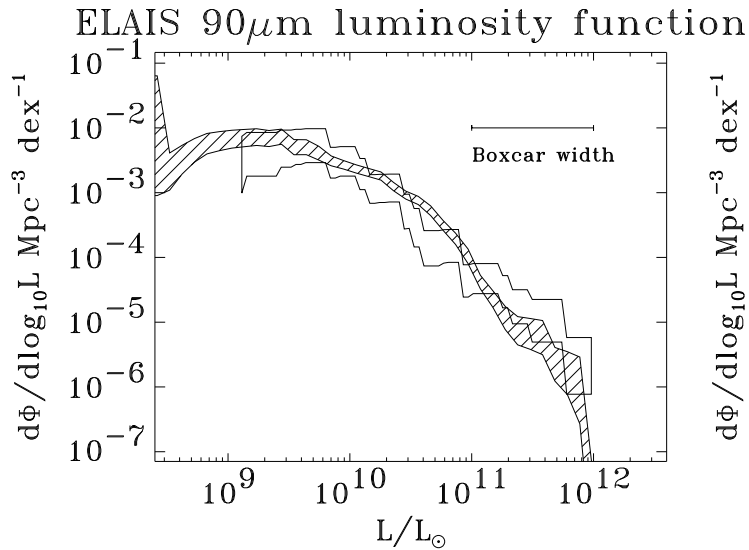}
\caption{\label{fig:lf_fa_and_pa}
Shaded area shows the $\pm1\sigma$ 
luminosity function constraint at $90\mu$m from the ELAIS
Final Analysis of the northern ELAIS fields. 
Pure luminosity evolution of $(1+z)^3$ is assumed, though this has
only a small effect on the derived luminosity function except at the
highest luminosities. 
Luminosities are
converted to bolometric luminosities 
assuming $\nu L_\nu=$constant. At each luminosity, the objects with
luminosities $\pm0.5$ dex are used in the construction of the
luminosity function, which is indicated on the diagram as the boxcar
width. Also plotted as the unshaded bound area is the $90\mu$m ELAIS
luminosity function from the Preliminary Analysis of the S1
field, which used the same boxcar width. Note the excellent
agreement between the Preliminary and Final Analyses, and the much
tighter constraint from the Final Analysis. 
}
\end{figure}

\begin{figure}
\centering
\ForceWidth{5in}
\hSlide{-1.8cm}
\vspace*{-1cm}
\BoxedEPSF{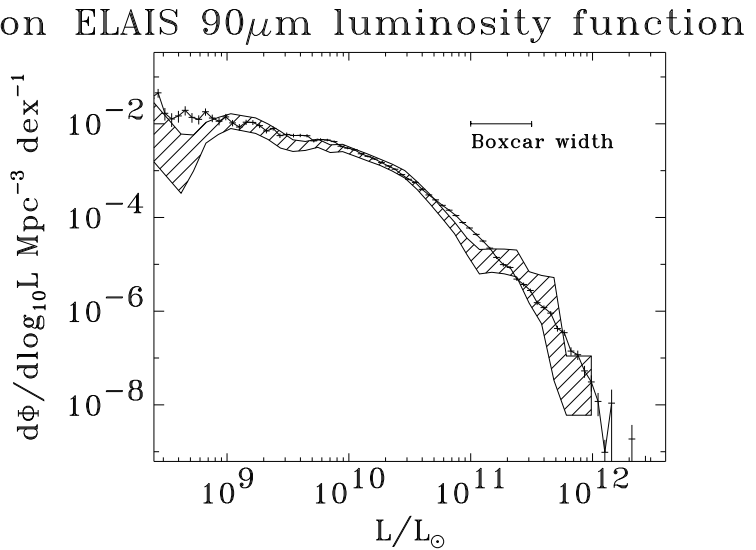}
\caption{\label{fig:lf_fa_and_pscz}
Shaded area shows the $\pm1\sigma$ 
luminosity function at $90\mu$m from the ELAIS
Final Analysis of the northern ELAIS fields, with a boxcar width of
$\pm0.25$dex (c.f. figure \ref{fig:lf_fa_and_pa}). 
Pure luminosity evolution of $(1+z)^3$ is assumed. 
Luminosities are
converted to bolometric luminosities 
assuming $\nu L_\nu=$constant. Also plotted is the predicted $90\mu$m
local luminosity function from the PSC-z survey (Serjeant \& Harrison
2004). Note the good agreement of the ELAIS luminosity function
with this prediction. 
}
\end{figure}

\section{Results}\label{sec:results}
A luminosity function can be constructed for any sample {\it provided}
that the selection function can be accurately stated, and that there
are no populations of objects which are undetectable at {\it any}
redshift. 
For a single flux-limited sample, the number density in a given
luminosity bin is given by 
\begin{equation}
\Phi = \Sigma V^{-1}_{{\rm max }, i}
\end{equation}
with an associated RMS error of 
\begin{equation}
\Delta\Phi = \sqrt{\Sigma V^{-2}_{{\rm max }, i}}
\end{equation}
where the sums are taken over the objects in the bin. The maximum volume
$V_{\rm max}$ for each object is the comoving volume enclosed by the
maximum redshift at which such an object is detectable, given the
flux limit. 

This procedure can easily be generalised for the presence of more
complicated selection criteria, including the multiwavelength flux
limits used in this paper. For example, if the selection function
is a function of redshift, then this function must be used to weight
the differential volume elements, which are then integrated to
calculate a weighted $V_{\rm max}$ for each object. 
Paper IV (Serjeant et al. 2001) 
discusses in more detail the formalism for generating 
$1/V_{\rm max}$ luminosity functions in the presence of complicated
selection functions. 
In this paper we largely restrict our discussion to
the construction of the selection function, and refer the reader to
paper IV for its application to the luminosity function. 
The only modification of this approach of Serjeant et al. (2001) to
this paper is in the 
consideration of bins containing only one galaxy. In this paper
these bins (and only these bins) are treated as having a $1\sigma$
range of $(0.18-3.3)V_{\rm max}^{-1}$ (rather than $V_{\rm
max}^{-1}\pm V_{\rm max}^{-1}$), 
corresponding to the $\pm1\sigma$ likelihood 
limits of an underlying Poisson
distribution with one detection (Gehrels 1986).

The optical completeness correction is the largest systematic
uncertainty in our selection function. To estimate this, we plot in
figure \ref{fig:optical_selection_function} the fraction of galaxies
with spectroscopic redshifts, as 
a function of apparent magnitude. Note that this is a {\it
differential} plot, i.e. at a magnitude $m$ we only consider galaxies
with magnitudes $m\pm1$. We found that our luminosity function
is robust to changes in the derivation of this optical
completeness, such as considering galaxies with magnitudes $m\pm1.5$
or $m\pm2$ instead of $m\pm1$, in that any systematic 
changes to the derived luminosity function are very much smaller than
the $\pm1\sigma$ random errors. 

\begin{figure}
\centering
\ForceWidth{5in}
\hSlide{-1.8cm}
\vspace*{-1cm}
\BoxedEPSF{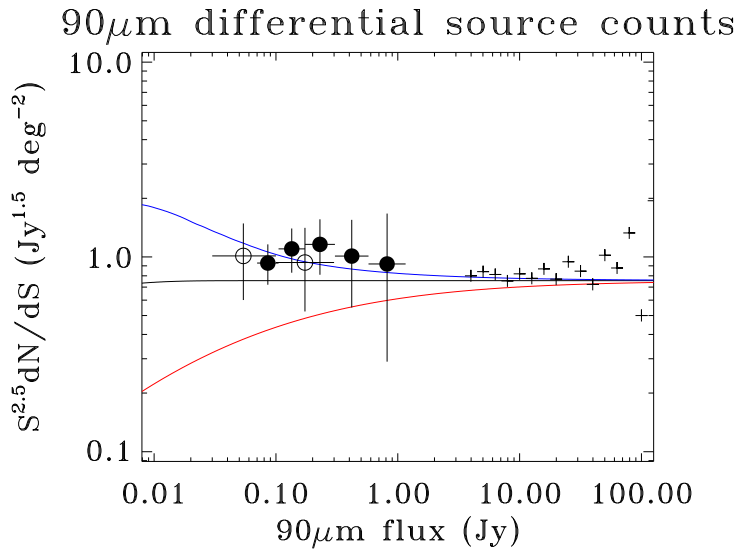}
\caption{\label{fig:counts}
Source counts from ELAIS (filled circles, H\'{e}raudeau et al. 2004), 
the Lockman Hole (open circles Rodighiero et al. 2003), and local
counts estimated from IRAS (crosses, Serjeant \& Harrison 2004). Also
overplotted are three model pure evolution curves of $(1+z)^\alpha$,
where $\alpha=0$ (bottom curve, no evolution), $\alpha=3$ (central
curve) and $\alpha=4$ (upper curve). All evolution is assumed to have
stopped in these models at $z=2$. Note the clear deviation of the
source counts from the no-evolution case. 
}
\end{figure}

To have an optical identification, each galaxy must either have a
bright optical ID ($B<18.5$) or be detected at $15\mu$m, $7\mu$m or in
the radio. In the case of $15\mu$m, $7\mu$m and radio detections, the
galaxy can be identified to fainter optical magnitudes ($B<21$), since
the greater astrometric accuracy of these catalogues makes reliable
identifications possible to fainter magnitudes. 
The probability of a given galaxy having an identification is therefore 
given by 
\begin{equation}
P_{\rm ID}=1-(1-p_{\rm opt})(1-p_{15\mu\rm m})(1-p_{7\mu\rm
m})(1-p_{\rm rad})
\end{equation}
Here $p_{\rm opt}$ is the probability that this galaxy has a bright
optical ID, and similarly for $15\mu$m, $7\mu$m and radio. For a
spectroscopically-complete sample, $p_{\rm opt}$ is 
$1$ at redshifts where the galaxy would have $B<18.5$, and $0$
at higher redshifts. In our case, we use the optical completeness
curve of figure \ref{fig:optical_selection_function} for the $B\leq18.5$
range. Similarly, $p_{15\mu\rm 
m}$ incorporates both the $15\mu$m completeness (e.g. Gruppioni et
al. 2003) and the optical completeness from figure
\ref{fig:optical_selection_function}, i.e.  
$p_{15\mu\rm m}=c_{15\mu\rm m}\times c_{\rm opt}$, where 
$c_{15\mu\rm m}$ is the $15\mu$m completeness and $c_{\rm opt}$ is the
completeness of optically-faint galaxies 
in the optical. Analogous optical completeness
corrections were made for $7\mu$m and radio identifications. 
We calculate $P_{\rm ID}(z)$ for every galaxy in our sample, and use
it to weight the volume elements in the calculation of the total 
accessible volume $V_{\rm max}$ (Avni \& Bahcall 1983) for each
galaxy, following the procedure in paper IV.  

What is the best way of representing the luminosity function? One can
make parametric fits, but these are subject to the assumptions in the
parameterisation; alternatively, 
one can directly bin the luminosity function, but
this loses information. 
Dunlop \& Peacock (1990) used a `free-form' fitting technique in 
their analysis of the 
luminosity function of radio-loud active
galaxies. 
In this methodology, a selection of high-order polynomials are fitted
to the data. The arbitrariness and variation of the fitting functions
partly avoids the issue of assumptions in the functional form. 
Divergence in the fits indicates lack of constraint, but unfortunately
convergence does not necessarily imply constraint. Also, the range
spanned by the models does not correspond to e.g. a $\pm1\sigma$
constraint. In this paper we adopt an alternative approach, similar to
that used in Serjeant et al. (2001): instead of binning the data into
discrete bins, we calculate the luminosity function at {\it every}
luminosity $L$ in a bin of logarithmic width $\pm\delta\log_{10}L$. 
This has the advantage of returning a $\pm1\sigma$ constraint at any
luminosity, while explicitly 
controlling the balance between resolution (in luminosity) and loss of
information with the choice of the boxcar size. 

Figure \ref{fig:lf_fa_and_pa} 
shows the
$90\mu$m $1/V_{\rm max}$ luminosity function derived in this way. We
compute the luminosity function in a logarithmic bin
of width $\pm0.5$ dex. 
Figure \ref{fig:lf_fa_and_pa} also gives a
comparison with the ELAIS Preliminary Analysis luminosity function
from the ELAIS S1 field (Serjeant et al. 2001), and the agreement is
excellent. However, a disadvantage of such a wide binning of the data
is that there are effectively very few independent data points
(e.g. $4$ for a data set spanning four decades in
luminosity). This was necessary for the small spectroscopic sample
($37$) in paper IV, but the current sample of $151$ is large enough to
benefit from finer binning. 
Therefore in figure \ref{fig:lf_fa_and_pscz} we plot the
$90\mu$m luminosity function for a 
$\pm0.25$ dex binning of the data. This figure also plots a
comparision with the predicted $90\mu$m local luminosity function from
the PSC-z survey (Serjeant \& Harrison 2004), estimated from the
IRAS colours of each of the $15411$ PSC-z galaxies. 
In both figure  \ref{fig:lf_fa_and_pa} and figure
\ref{fig:lf_fa_and_pscz}, the population is assumed to evolve with
pure luminosity evolution at a rate of $(1+z)^3$. This assumption does
not affect the luminosity function determined from the data except at
the brightest end (figure \ref{fig:luminosity_redshift}). 
The luminosity function is tabulated in table \ref{tab:lf}. 

\begin{table}
\begin{tabular}{lll}
Luminosity bin & $\Phi$ & $\Delta\Phi$\\
$\log_{10}(L/L_\odot)$ & Mpc$^{-3}$ dex$^{-1}$ & Mpc$^{-3}$ dex$^{-1}$\\
\hline
$8.625\pm0.25$ & $1.86\times10^{-3}$ & $1.84\times10^{-3}$ \\
$8.75\pm0.25$ & $4.1\times10^{-3}$ & $2.62\times10^{-3}$ \\
$8.875\pm0.25$ & $8.4\times10^{-3}$ & $3.64\times10^{-3}$ \\
$9.\pm0.25$ & $1.14\times10^{-2}$ & $4.13\times10^{-3}$ \\
$9.125\pm0.25$ & $1.1\times10^{-2}$ & $3.89\times10^{-3}$ \\
$9.25\pm0.25$ & $9.38\times10^{-3}$ & $3.41\times10^{-3}$ \\
$9.375\pm0.25$ & $6.3\times10^{-3}$ & $2.35\times10^{-3}$ \\
$9.5\pm0.25$ & $4.07\times10^{-3}$ & $1.3\times10^{-3}$ \\
$9.625\pm0.25$ & $3.47\times10^{-3}$ & $7.84\times10^{-4}$ \\
$9.75\pm0.25$ & $3.88\times10^{-3}$ & $7.68\times10^{-4}$ \\
$9.875\pm0.25$ & $3.02\times10^{-3}$ & $5.57\times10^{-4}$ \\
$10.\pm0.25$ & $2.75\times10^{-3}$ & $4.73\times10^{-4}$ \\
$10.125\pm0.25$ & $2.06\times10^{-3}$ & $3.49\times10^{-4}$ \\
$10.25\pm0.25$ & $1.53\times10^{-3}$ & $2.38\times10^{-4}$ \\
$10.375\pm0.25$ & $1.09\times10^{-3}$ & $1.87\times10^{-4}$ \\
$10.5\pm0.25$ & $6.66\times10^{-4}$ & $1.24\times10^{-4}$ \\
$10.625\pm0.25$ & $3.16\times10^{-4}$ & $6.8\times10^{-5}$ \\
$10.75\pm0.25$ & $1.41\times10^{-4}$ & $3.45\times10^{-5}$ \\
$10.875\pm0.25$ & $5.85\times10^{-5}$ & $1.66\times10^{-5}$ \\
$11.\pm0.25$ & $2.14\times10^{-5}$ & $9.15\times10^{-6}$ \\
$11.125\pm0.25$ & $1.38\times10^{-5}$ & $7.27\times10^{-6}$ \\
$11.25\pm0.25$ & $1.38\times10^{-5}$ & $7.4\times10^{-6}$ \\
$11.375\pm0.25$ & $1.28\times10^{-5}$ & $7.39\times10^{-6}$ \\
$11.5\pm0.25$ & $3.98\times10^{-6}$ & $2.71\times10^{-6}$ \\
$11.625\pm0.25$ & $2.92\times10^{-6}$ & $2.61\times10^{-6}$ \\
$11.75\pm0.25$ & $1.54\times10^{-7}$ & $1.53\times10^{-7}$ \\
$11.875\pm0.25$ & $3.34\times10^{-8}$ & $^{+7.68}_{-2.74}\times10^{-8}$ \\
\end{tabular}
\caption{\label{tab:lf}
Tabulated luminosity function with boxcar width $\pm0.25$ dex, as
plotted in figure \ref{fig:lf_fa_and_pscz}.}
\end{table}

\section{Discussion and Conclusions}\label{sec:discussion} 



Using differential volume elements weighted to the selection function
(section \ref{sec:results}), 
the $\langle V/V_{\rm max}\rangle$ distributions hint at the
existence of evolution in this population, but only at the
$\sim0.6\sigma$ level: assuming a no-evolution
model yields $\langle V/V_{\rm max}\rangle = 0.523\pm 0.029$, while
assuming $(1+z)^3$ pure luminosity evolution yields 
$\langle V/V_{\rm max}\rangle = 0.482\pm 0.029$. With this test, 
deviation from $1/2$ is evidence for evolution (Avni \& Bahcall
1983). However, this statistic is not the most efficient at detecting
evolution. 

Figure \ref{fig:lf_fa_and_pscz} demonstrates that the ELAIS $90\mu$m
luminosity function is consistent with the prediction from the local
IRAS population by Serjeant \& Harrison (2004). If we {\it assume}
that this prediction is correct, then we can use the source counts to
constrain the strength of the evolution. Figure \ref{fig:counts} plots
the ELAIS $90\mu$m Final Analysis source counts (H\'{e}raudeau et
al. 2004), together with the interpolated $90\mu$m IRAS counts from
Serjeant \& Harrison (2004). We also plot the predicted source counts
for the predicted $90\mu$m luminosity function with $(1+z)^\alpha$
pure luminosity evolution to $z=2$, with $\alpha=0,3,4$. Clearly, the
source counts are a strong discriminant of the strength of the
evolution. Provided the local luminosity function is of this form, the
constraint on the evolution is $2.4<\alpha<4.4$ at $68\%$ confidence,
or $1.3<\alpha<4.6$ at $95\%$ confidence. 
Is $z=2$ the most appropriate redshift to halt the evolution?
There is evidence that the evolution in the populations at
$15\mu$m stops at $z\sim1$ (Franceschini et al. 2001, Chary \& Elbaz
2001, Xu et al. 2003). 
Decreasing the maximum
redshift to which pure luminosity evolution extends from $z=2$ to
$z=1$ slightly alters our constraints to $2.4<\alpha<4.6$ at $68\%$
confidence, or $1.3<\alpha<4.8$ at $95\%$ confidence. This evolution
is consistent with that derived by us in paper IV. Note that pure
density evolution (PDE) can already be 
excluded as it overpredicts the sub-mJy
radio source counts (Rowan-Robinson et al. 1993), and we do not
consider PDE models further in this paper; however, it is worth
noting that the far-infrared data alone is {\it not} sufficient to
rule out pure density evolution. 

This evolution rate is in good agreement with that derived from
optically-selected galaxy samples (e.g. SLOAN, Glazebrook et
al. 2003). Our data is consistent with the obscured star formation
history and the corresponding unobscured history both evolving at
similar rates, although the data at $90\mu$m also permits stronger
evolution than that determined from the optically-selected
samples. This 
is in agreement with the determinations from the sub-mJy 
radio population (e.g. Haarsma et al. 2000, Gruppioni et al. 2001)
which are sensitive to both the obscured and unobscured high-mass 
star formation. However, the apparent plateau in the cosmic star
formation history above $z=2$ indicated by both optically-selected
galaxies (e.g. Thompson 2003) and in the sub-mm (e.g. Hughes et
al. 1998, Scott et al. 2002, Fox et al. 2002) is difficult to
reconcile with the current comoving mass density in stars,
$\Omega_\ast$ (e.g. Blain et al. 1999, Serjeant \& Harrison 2004). For
the infrared-luminous population, this may be in part due to
contamination of the far-infrared flux by cirrus heated by the
interstellar radiation field (Efstathiou \& Rowan-Robinson 2003,
Kaviani et al. 2003). Alternatively, the initial mass function at high
redshift may be skewed to high masses (e.g. Larson 1998, Franceschini
et al. 2001, Serjeant \& Harrison 2004). 

The {\it Spitzer} satellite has already launched successfully, and the
prospects for improving on this analysis in the near future are
excellent. The SWIRE survey (Lonsdale et al. 2003, 2004) will survey $70$
square degrees including all the major ELAIS fields, 
deeper at $70\mu$m than the ELAIS survey at $90\mu$m. The
source counts will be a powerful discriminant of the evolution, but
more importantly the existence of large spectroscopic redshift surveys
in the SWIRE fields (Rowan-Robinson et al. 2004) will be a powerful
and immediate tool for the exploitation of this important legacy data
set. 

\section{Acknowledgements}
We would like to thank the support staff at the Guillermo Haro
observatory, the Isaac Newton Group, and Kitt Peak for their help
during the several observing runs in which we collected data for this
paper. We would also like to thank the anonymous
referee for useful comments. 
This research was supported by PPARC grant PPA/G/0/2001/00116, and
Nuffield Foundation grant NAL/00529/G. 
ELAIS was also supported by EU TMR Network FMRX-CT96-0018. 
This research made use of the NASA/IPAC Extragalactic
Database (NED) which is operated by the Jet Propulsion Laboratory,
California Institute of Technology, under contract with the National
Aeronautics and Space Administration. 
This paper is based on observations with {\it ISO}, an ESA project
with instruments funded by ESA member states (especially the PI
countries: France, Germany, the Netherlands and the United Kingdom)
and with the participation of ISAS and NASA.

\end{document}